\documentclass[sigconf]{acmart}

\AtBeginDocument{%
  }

\setcopyright{acmlicensed}
\copyrightyear{2018}
\acmYear{2018}
\acmDOI{XXXXXXX.XXXXXXX}

\acmConference[Conference acronym 'XX]{Make sure to enter the correct
  conference title from your rights confirmation emai}{June 03--05,
  2018}{Woodstock, NY}
\acmISBN{978-1-4503-XXXX-X/18/06}

\usepackage{xspace}     
\usepackage{xpunctuate} 
\usepackage[nounderscore]{syntax}

\newcommand{\dageda}{DAG}

\newcommand{\eg}{{e.g.,}\xspace}
\newcommand{\etal}{{et~al\xperiod}\xspace}

\newcommand{\etc}{{etc\xperiod}\xspace}  

\newcommand{\myquote}[1]{\emph{``#1''}} 

\newcommand{\parHeading}[1]{\vspace{8px}\noindent{\textbf{#1}}} 

\usepackage{xargs}      
\usepackage{soul}       
\usepackage{color}      

\definecolor{LIGHTPINK}{RGB}{237,157,202}
\definecolor{LIGHTRED}{RGB}{210,121,121}
\definecolor{LIGHTORANGE}{RGB}{230,170,50}
\definecolor{LIGHTGOLD}{RGB}{210,194,121}
\definecolor{LIGHTGREEN}{RGB}{121,210,121}
\definecolor{LIGHTAQUA}{RGB}{121,206,210}
\definecolor{LIGHTBLUE}{RGB}{121,124,210}
\definecolor{LIGHTPURPLE}{RGB}{153,102,255}
\definecolor{RED}{RGB}{178,34,34}
\definecolor{GRAY}{RGB}{166,166,166}
\definecolor{WHITE}{RGB}{255,255,255}

\newif\ifshowcomments
\showcommentstrue

\newcommandx{\leilani}[2][1=] 
    {\setulcolor{LIGHTPURPLE}{\ul{#1}} \textcolor{LIGHTPURPLE}
    {[\textbf{Leilani:} #2]}}

\newcommandx{\junran}[2][1=] 
    {\setulcolor{LIGHTORANGE}{\ul{#1}} \textcolor{LIGHTORANGE}
    {[\textbf{Junran:} #2]}}
\newcommandx{\teanna}[2][1=] 
    {\setulcolor{LIGHTBLUE}{\ul{#1}} \textcolor{LIGHTBLUE}
    {[\textbf{Teanna:} #2]}}
\newcommandx{\shruti}[2][1=] 
    {\setulcolor{LIGHTPINK}{\ul{#1}} \textcolor{LIGHTPINK}
    {[\textbf{Shruti:} #2]}}
\newcommandx{\guest}[3][1=]
    {\setulcolor{LIGHTORANGE}{\ul{#1}} \textcolor{LIGHTORANGE} 
    {[\textbf{#2:} #3]}}

\newcommand{\dagGrammarTable}{
\begin{table*}[!tp]
\centering
\small
\caption[The reference instantiation of the typology: base states.]{The reference instantiation of the typology, as the two tables behind $f$ and $g$. $f$ is defined only up to arity two: these four multisets are every base state the grammar admits. $g$ is multi-valued where more than one framing is defensible; choosing among them is a render-time decision, not a production. Swapping these tables swaps the analytical vocabulary. Every analysis over three or more variables is a composite, given in \autoref{tab:dag-composites}. Examples are drawn from the dataset of Section~\ref{sec:usage}.}
\label{tab:dag-grammar}
\renewcommand{\arraystretch}{1.18}

\begin{tabular}{@{}llllp{2.15in}@{}}
\toprule
\textsc{multiset} & $k$ & \textsc{$f$: hypothesis} & \textsc{$g$: template} & \textsc{example claim} \\
\midrule
\texttt{\{Q\}}       & 1 & Value      & \texttt{M1}          & The distribution of worldwide gross. \\
\midrule
\texttt{\{Q,\,Q\}}   & 2 & Relation   & \texttt{M3}          & Production budget against worldwide gross. \\
\texttt{\{Q,\,C\}}   & 2 & Comparison & \texttt{CAT1 M2}     & Rotten Tomatoes rating across genre. \\
\texttt{\{Q,\,T\}}   & 2 & Trend      & \texttt{CH1}         & Worldwide gross over release year. \\
\bottomrule
\end{tabular}

\end{table*}
}

\newcommand{\dagCompositeTable}{
\begin{table*}[!tp]
\centering
\small
\caption[The reference instantiation of the typology: composites.]{The three composites of the reference instantiation, which extend the base states of \autoref{tab:dag-grammar} and are the only route to an analysis over three or more variables. Composites are not states and so have no image under $f$ or $g$. Each carries a well-formedness condition that is \emph{relational}, a property of how its sub-specs fit together that a single state cannot express, and each compiles to a qualitative layout over templates rather than to one template. The layouts subsume the multi-variable templates directly: a Condition over a Comparison yields the two-way \texttt{CAT2} crosstab, and a Compare over two Trends sharing a time axis yields the \texttt{CH2} dual-series chart. Examples are drawn from the dataset of Section~\ref{sec:usage}.}
\label{tab:dag-composites}
\renewcommand{\arraystretch}{1.18}

\begin{tabular}{@{}lp{1.55in}p{1.5in}p{1.35in}@{}}
\toprule
\textsc{composite} & \textsc{well-formed when} & \textsc{compiles to} & \textsc{example} \\
\midrule
$\text{Condition}(\textit{Spec}, z)$ &
$z$ is categorical. It may be left open and bound afterwards. &
A facet over $z$: small multiples or a colour encoding; the \texttt{CAT2} crosstab where the conditioned spec is a Comparison. &
Does the budget--gross relation hold across genres? \\
\addlinespace
$\text{Chain}(\textit{Spec}, \textit{Spec}, m)$ &
Both legs are Relations, both bind $m$, and their remaining ends differ. &
Two linked Relation panels, brushed on the shared $m$. &
Budget $\rightarrow$ vote count $\rightarrow$ worldwide gross. \\
\addlinespace
$\text{Compare}(\textit{Spec}, \dots)$ &
At least two legs, sharing a hypothesis type, pairwise distinct. &
Overlay on the shared axis for Relations and for Trends, the latter giving \texttt{CH2}; a slope chart for Comparisons; side-by-side otherwise. &
The box-office reading against the critical reading. \\
\bottomrule
\end{tabular}

\end{table*}
}

\newcommand{\dagArch}{
    \begin{figure*}[!tp]
        \centering
        \includegraphics[width=.7\textwidth]{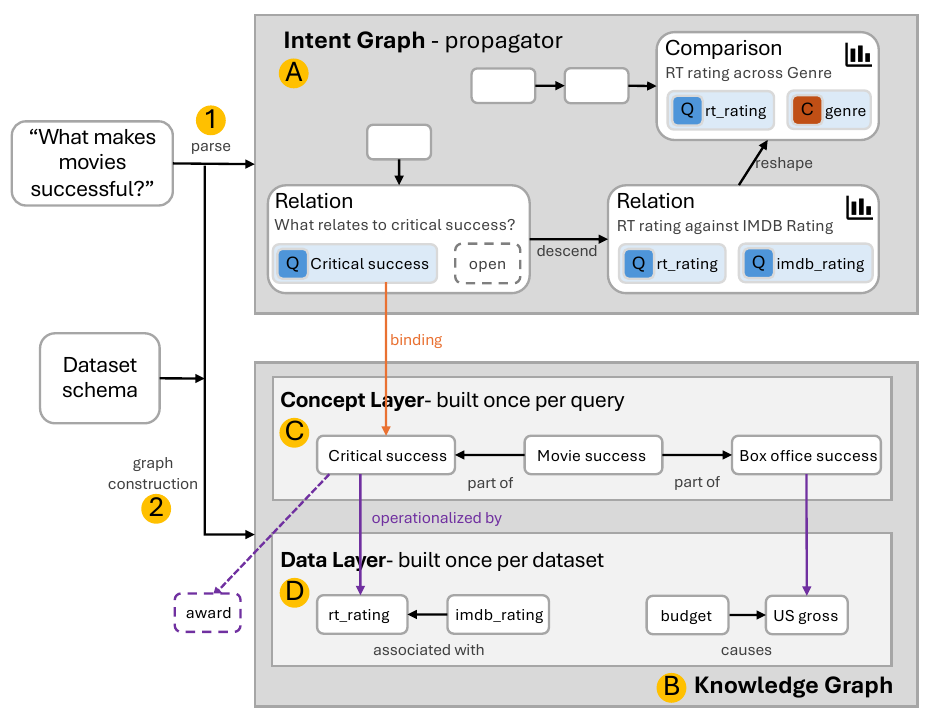}
        \caption[The \dageda{} architecture.]{The \dageda{} architecture, instantiated on \myquote{What makes movies successful?}
        The analyst's question and the dataset schema are the only inputs.
        \textcircled{\footnotesize 1} Parsing the question seeds the intent graph \textcircled{\footnotesize A} with the claims it could resolve into, and \textcircled{\footnotesize 2} graph construction builds the knowledge graph \textcircled{\footnotesize B} from the question and the schema together.
        In the intent graph, each node is a claim whose slots carry a type, measurable (Q) or categorical (C), and a binding that is open, an unresolved concept, or a set of columns; a chart icon marks a node whose required slots are grounded and which therefore renders to a dashboard.
        The two edge families appear here as one \emph{descend} step, which resolves \myquote{critical success} to the columns beneath it, and one \emph{reshape} step, which turns a relationship into a comparison by admitting a categorical slot.
        The knowledge graph splits into a concept layer \textcircled{\footnotesize C}, built once per question, and a data layer \textcircled{\footnotesize D}, built once per dataset, joined by \emph{operationalized by} bridge edges.
        The \emph{binding} edge is where the two structures meet: grounding a slot is a walk from the concept a slot holds to the columns that operationalize it.
        The dashed \myquote{award} concept has no column to operationalize it, which the system reports rather than silently dropping.}
        \Description{}
        \label{fig:dag-arch}
    \end{figure*}
}

\newcommand{\dagTeaser}{
    \begin{figure*}[t!]
        \centering
        \includegraphics[width=\textwidth]{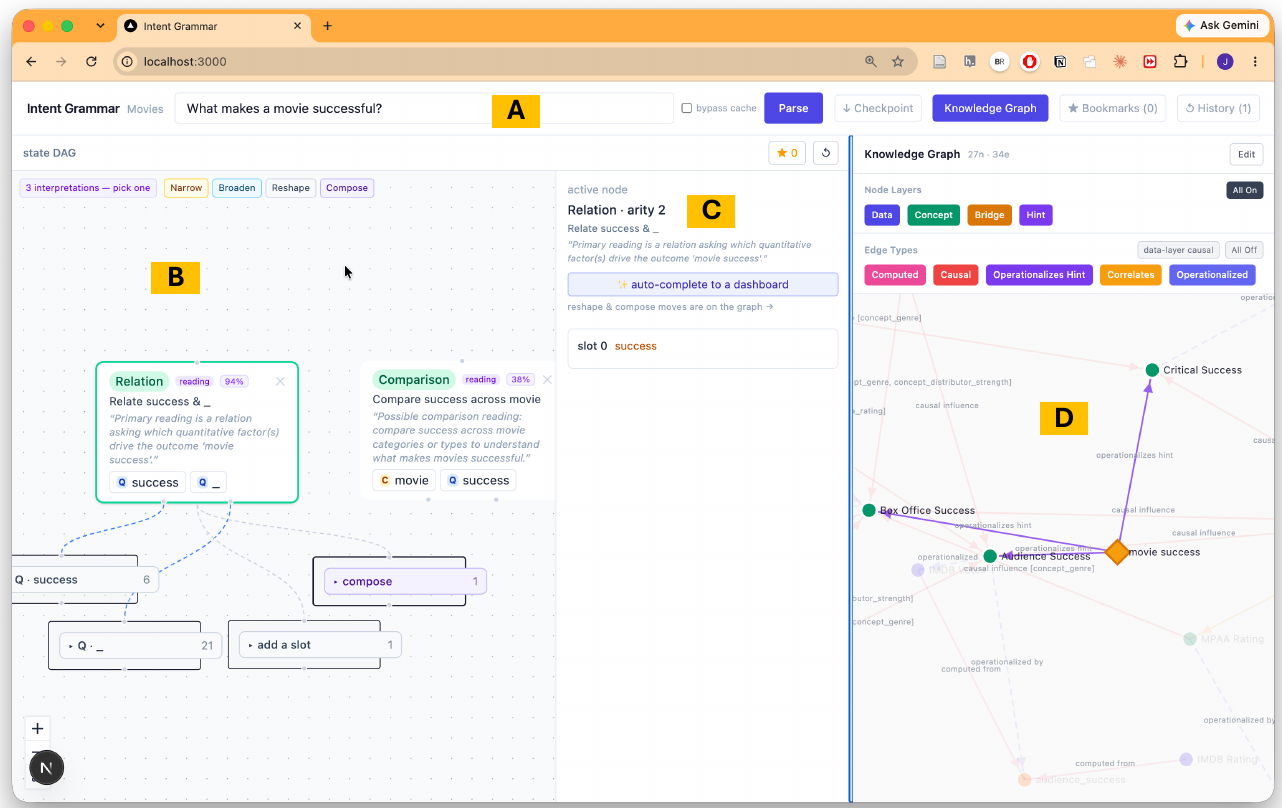}
        \caption[The \dageda{} interface.]{The \dageda{} interface, early in Anna's session.
        The question bar (A) holds the analytical question the session started from and the control that parses it.
        The state DAG canvas (B) shows the claims Anna has formed, one node per candidate analysis.
        Here it holds the interpretations the parser returned, each labeled with the claim it makes, a natural-language rendering of what it would ask, the confidence that it is the intended reading, and its slots as chips.
        The moves available from each node hang beneath it, grouped by the state they would produce and collapsed to a count, so that the shape of what is still open is visible without enumerating it; the toolbar names the four families of move the grammar admits.
        The active node panel (C) describes the selected claim in full, lists its slots, and offers to complete it to a dashboard once its required slots are grounded.
        The knowledge graph (D) shows the concept layer and the data layer joined by bridge edges, with filters for which layers and edge types to display.}
        \Description{}
        \label{fig:dag-teaser}
    \end{figure*}
}

\newcommand{\dagStoryboard}{
    \begin{figure*}[tp!]
        \centering
        \includegraphics[width=\textwidth,pagebox=cropbox]{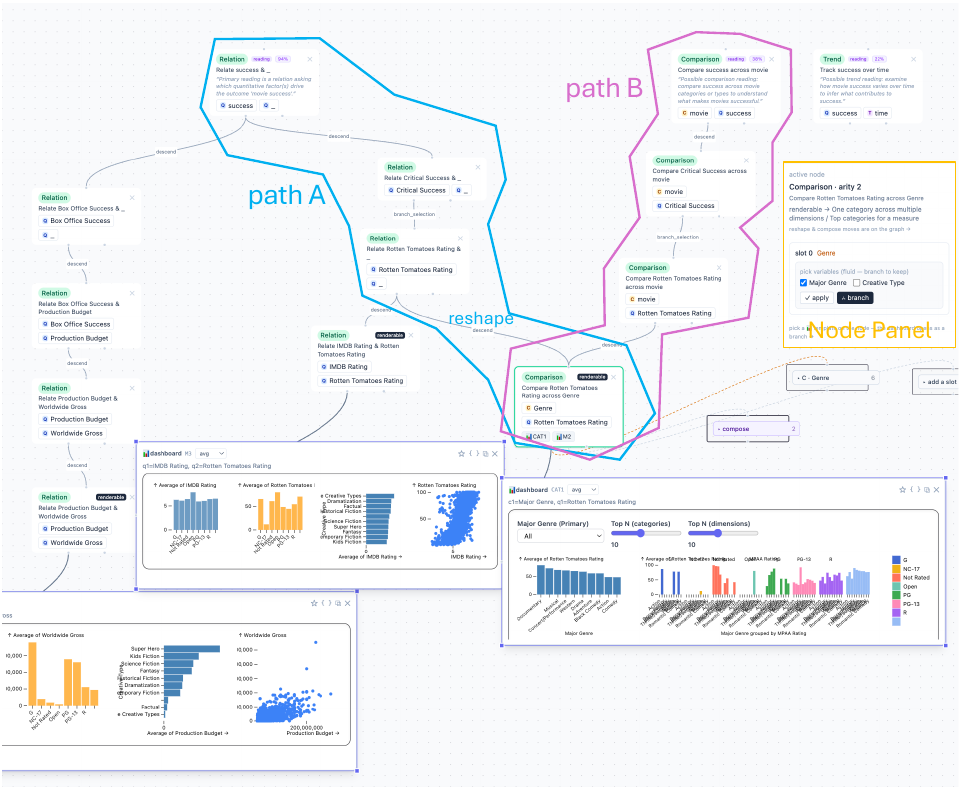}
        \caption[Anna's intent graph at the end of her session.]{Anna's intent graph at the end of her session, with the dashboards she has open and the node she has selected.
        Across the top are the three readings the parser returned for \myquote{What makes a movie successful?}
        Two paths A and B are outlined. 
        Each outline encloses the claims Anna formed while following that reading, and the two enclose one card in common, the comparison of Rotten Tomatoes rating across genre that both readings resolve to.
        The move marked \emph{reshape} is the one that changes the shape of the claim as it binds, which turns a relation into a comparison. 
        The node panel at the right describes the selected claim
        A node is marked \emph{renderable} once its slots reach data and it can be completed to a dashboard.
        }
        \Description{}
        \label{fig:dag-storyboard}
    \end{figure*}
}
\begin{document}

\title{Intent Graph: Navigating the Analytical Reasoning Space for Exploratory Data Analysis}

\author{Junran Yang}
\email{junran@cs.washington.edu}
\orcid{}
\affiliation{%
  \institution{University of Washington}
  \city{Seattle}
  \state{Washington}
  \country{USA}
}

\author{Shruti Badrish}
\email{shrutib@uw.edu}
\orcid{}
\affiliation{%
  \institution{University of Washington}
  \city{Seattle}
  \state{Washington}
  \country{USA}
}

\author{Teanna Barrett}
\email{tb23@cs.washington.edu}
\orcid{}
\affiliation{%
  \institution{University of Washington}
  \city{Seattle}
  \state{Washington}
  \country{USA}
}

\author{Leilani Battle}
\email{leibatt@cs.washington.edu}
\orcid{}
\affiliation{%
  \institution{University of Washington}
  \city{Seattle}
  \state{Washington}
  \country{USA}
}
\renewcommand{\shortauthors}{Trovato et al.}

\begin{abstract}


Exploratory data analysis (EDA) is rarely open-ended in practice: analysts work from high-level domain questions toward the concrete analyses that can answer them, prioritizing directions with domain knowledge and prior hypotheses.
Large language models (LLMs) can supply such knowledge, but their responses are unstructured, leaving analysts no way to see what has been explored, what is missing, or why one direction was chosen over another.
We present \dageda{}, a system that lets analysts and an LLM co-navigate the space of possible analyses through two linked structures.
An \emph{intent graph}, governed by a grammar of analytical intent, decomposes an ambiguous natural-language question into progressively concrete analysis tasks, keeping alternative framings open and letting analysts branch, backtrack, and compare paths.
A multi-layered \emph{knowledge graph} externalizes the LLM's domain knowledge, linking domain concepts to the dataset variables that can measure them, so analysts can inspect and contest how their question is grounded in the data.
Both graphs are constructed from only the dataset and the analyst's question, and the analyses the analyst reaches are rendered as interactive dashboards.
We illustrate the system through a usage scenario and describe a user study design for examining whether the system scaffold analysts' reasoning and navigation.

\end{abstract}



\received{20 February 2007}
\received[revised]{12 March 2009}
\received[accepted]{5 June 2009}

\maketitle



\section{Introduction}
The traditional view of exploratory data analysis (EDA) was introduced by Tukey as an open-ended, bottom-up process of finding interesting patterns in a dataset, operationalized by domain-agnostic approaches such as data profiling (\eg{} summarize distributions, flag anomalies, enumerate correlations, \etc{}). 
This framing has shaped how many EDA tools are built: treating variables and their combinations as equally worth examining~\cite{wongsuphasawat_voyager_2016, trifacta_wrangler_2017}, or passively surfacing variables of interest through statistics~\cite{vartak_seedb_2015}.
Due to the opportunistic nature of EDA, analysts face two interleaving challenges.
First, the space of possible variable combinations is vast.
Without a principled basis for prioritization, analysts must make ad hoc decisions about where to look. 
Second, there is no natural stopping condition since EDA has no formal criterion of completion. 
Both challenges are compounded by the reality that analysts operate under limited time, resulting in them moving on to other tasks before feeling ready~\cite{wongsuphasawat2019goals}. 

In reality, analysts do not treat all attribute combinations as equally worth exploring. They prioritize based on domain knowledge, prior hypotheses, and what they already believe matters~\cite{wongsuphasawat2019goals, lam2018bridging}. 
This is not a deviation from good analytical practice, but a necessary and plausible approach to an otherwise intractable search problem. 
In practice, the EDA process is often guided by high-level intents, motivated by the context in which the analysis is embedded, such as business requirements, stakeholder questions, or research goals~\cite{alspaugh2018futzing}. 
EDA activities exist along a spectrum from exploratory to directed, at the far end of which executions can be clearly specified by top-down planning~\cite{alspaugh2018futzing}. 
These intents do not specify the analysis in detail, but they shape which directions are worth pursuing and which findings would be considered relevant. EDA, in this light, is less about open-ended discovery and more about goal-directed analytical reasoning: working from a high-level domain question toward the concrete analyses that can answer it.

Pre-LLM tools automate steps that are mechanical and repeatable to reduce manual implementation burden. Domain knowledge is still required from the analyst to interpret results and prioritize directions. Deciding what to look for, and why, remains unsupported. Analysts' depth and breadth of exploration are constrained by the time required to manually construct and inspect enough combinations.
LLMs change what is possible by surfacing which variables are likely to matter, which relationships are theoretically plausible, and which analytical directions are worth pursuing given a domain context. 
However, LLM responses are unstructured and not navigable: they offer no way to enumerate what has been explored, identify what is missing, or explain why one direction is more appropriate than another given the data available. 
Critically, the analyst has no way to see into the LLM's reasoning, contest its assumptions, or steer the exploration with confidence. 
This motivates our research question: \textbf{How can we enable users and LLMs to co-navigate the space of possible analyses during EDA in a transparent way?}
We identify two intertwined challenges that together constitute this problem.

\textbf{RQ1: Scaffolding and Navigation: } \textit{How can systems scaffold the analytical reasoning process and help analysts navigate from ambiguous, open-ended goals toward concrete data operations without over-constraining their exploration?} 
Analytical questions are ambiguous at several levels at once: in how the question is framed (exploratory vs. confirmatory), in what the terms mean (concepts vs. variables), and in which of many valid analytical directions to pursue. 
This ambiguity is a natural property of open-ended questions, so scaffolding must resolve it without prematurely eliminating alternatives the analyst might find inspiring. 
At the same time, the reasoning process itself is not a single decision but a sequence of choices across a space of possible framings, variables, and operations, and it is nonlinear: analysts branch into parallel framings, backtrack when a direction proves unfruitful, and reframe their question in light of what they discover. Without support, analysts lose track of where they are, how their current thread relates to others, and how to recover a coherent picture of their own reasoning process. 
Scaffolding this process requires defining what analytical moves are valid and in what order. 

\textbf{RQ2: Shared Representation of Analytical Intent: } \textit{How can systems establish a shared representation of analytical intent that remains meaningful to human analysts, LLMs, and executable code while preserving the ambiguity and flexibility that exploration requires?} 
Humans reason in domain concepts, LLMs operate on natural language, and code operates on data variables. For a system to meaningfully augment rather than replace analytical reasoning~\cite{heer2019agency}, all three must stay aligned around a common understanding of what the analyst is trying to do, yet each operates in a fundamentally different vocabulary. A shared representation must bridge all three to balance competing demands: it must be concrete enough for the system to act on and specific enough for analysts to have agency over their intent, yet flexible enough to hold multiple interpretations open until the analyst is ready to commit. Resolving ambiguity too early forecloses valid analytical directions; leaving it unresolved too long prevents the system from providing meaningful guidance. 

We address these challenges through a system that bridges the gulf between high-level analytical goals and low-level data operations by establishing shared context between the user and the LLM.
We propose an intent grammar, which formally constrains how intents instantiate, propagate across parameters, and evolve. 
Complementing the grammar, an NL parsing component translates open-ended analytical questions into the initial intent specification at the entry point to be propagated. 
To let analysts traverse the structure, maintain awareness of where they are in the process, and revisit prior branches, the system interface extends the capability of the grammar, making the grammar rules navigable and interactive. 
The intent specification serves as this shared language for goals, while the knowledge graph externalizes LLM-generated domain knowledge into a form the analyst can inspect and agree with, bridging the gap between what the system infers and what the analyst understands.

These two challenges are deeply interdependent. Shared representation is what makes scaffolded navigation meaningful: analysts are not merely traversing an abstract graph but moving through a conceptual space they and the system mutually understand. Scaffolding is what makes shared representation actionable: the representation is not a static snapshot but becomes progressively more concrete as the analyst makes choices, with each grounded parameter narrowing the space of what comes next. 
The interface operationalizes both challenges together: the intent DAG makes the scaffolded reasoning process traversable, while the knowledge graph visualization makes the shared representation legible. Neither component alone addresses both challenges. It is their combination that constitutes our system contribution.

We demonstrate \dageda{} through a walkthrough of an analyst working from an ambiguous question to a grounded dashboard (\autoref{sec:usage}).
What this chapter reports is the design and the rationale behind it; an empirical evaluation of how analysts use the two graphs remains to be carried out.

In this chapter, we make the following contributions:
\begin{itemize}
    \item \textbf{A graph-based representation of analytical intent} that scaffolds the analyst's intent and reasoning process for conducting exploratory data analysis (EDA) while authoring dashboards (\autoref{sec:dag-intent-graph}). The formalism consists of multiple granularities, from a high-level NL question to a concrete set of analysis tasks, supporting flexible navigation and exploration style.

    \item \textbf{A multi-layered knowledge graph} that informs the construction of the intent graph (\autoref{sec:dag-kg}). It consists of two layers. The concept layer contains domain concepts that guide the decomposition of high-level questions into investigable sub-questions. The data layer, containing causal relationships of data variables existing in the provided dataset, grounds abstract concepts to measurable data that can be visualized based on concrete analytical tasks.

    \item \textbf{LLM-based methods that construct both graphs from minimal context}, namely the dataset and the analytical question (\autoref{sec:impl-composition}). The framework consists of a bottom-up approach that groups data columns into semantically meaningful domain concepts and a top-down approach that extracts domain concepts based on the question, so that the knowledge graph is generated and the intent graph construction steered without a curated ontology for the domain.

    \item \textbf{An interface that integrates NL queries with intent graph manipulation} (\autoref{sec:impl-interface}), supporting iterative refinement of EDA intent and dashboard creation, and generating interactive dashboards from the terminal analyses the analyst reaches. Closing the loop, so that the system reads the analyst's interactions with those dashboards to update the intent graph, is a direction we outline as future work.

\end{itemize} 

\section{Related Work}
\label{sec:relatedwork}

\parHeading{Characterizing analytical intent.}
Intent is commonly described as the translation of a high-level question into low-level actions~\cite{brehmer2013multi-level}, and has been formalized at several granularities: classes of analysis goals derived from design studies~\cite{lam2018bridging}, low-level task taxonomies~\cite{amar2005low, narechania2020nl4dv}, and narrower operationalizations as attribute-and-filter selections~\cite{lee2021lux}, an ``interestingness'' metric over analytic actions~\cite{lee2021deconstructing}, or prompts to be triaged by a data-driven system~\cite{vaithilingam2025semantic, lipman2024linx}.
Pyxis~\cite{battle2024what} is closest to our stance, placing insight on two axes, analytic and domain knowledge, so that an analyst's beliefs about the domain are as constitutive of an insight as the data relationships they infer; hypotheses play a corresponding role in models of knowledge generation~\cite{sacha2014knowledge}.
A related line formalizes intent as a grammar rather than a taxonomy.
Suh~\etal{}~\cite{suh2022grammar} contribute a hypothesis space, containment, and grounding, but their grammar is \emph{descriptive}, enumerating every parseable hypothesis; ours is \emph{prescriptive}, constraining the semantic gaps a language model fills rather than closing a language.
For the shapes a grounded claim can take we adopt Choi~\etal{}'s recursive typology~\cite{choi2019concept}, which shares its goal vocabulary with Lam~\etal{}~\cite{lam2018bridging} but carries no notion of sequence; composing its types into a graph reintroduces ordering and makes it dynamic, since the shortest route to an unvisited analysis is recomputed as the analyst descends.
Outside visualization, work on search and conversational intent likewise treats intent as inferred and progressively refined rather than stated once~\cite{choi2025bloomintent, mei2025interquest}.

\parHeading{Dashboard recommendation.}
Dashboard automation is well surveyed~\cite{zeng2024systematic} and divides into approaches that encode designer knowledge as intents, templates, or mined rules~\cite{pandey2023medley, srinivasan2023bolt, epperson2023declarative, lin2024dminer} and approaches that learn composition from data~\cite{chen2022pi2, wu2022multivision, deng_dashbot_2022, purich2025adaptive}.
Medley~\cite{pandey2023medley} derives high-level dashboard intents from over 200 dashboards, and Bolt~\cite{srinivasan2023bolt} maps natural language onto a taxonomy of five objectives, whose authors note that static taxonomies cannot flexibly translate an utterance into low-level tasks.
Both families share two limits: the set of intents a system can implement stays small, and both concentrate on the analytic dimension of intent while leaving its domain knowledge dimension implicit.
\dageda{} answers the first by generating the decomposition per dataset and per question instead of fitting an utterance to a fixed taxonomy, and the second by constructing an explicit conceptual knowledge graph alongside it.
Work on coordinating multiple views and recording how they were reached~\cite{bavoil2005vistrails, silva2010using}, on characterizing dashboard design and use~\cite{Sarkaya2019WhatDW, Bach2022DashboardDP, srinivasan2024dashboard, tory_finding_2023}, on mapping data to effective encodings~\cite{moritz2018formalizing, hu2019vizml, luo2018deepeye} and extracting and sequencing data facts~\cite{wang_datashot_2020, shi_calliope_2021, ding2019quickinsights, Srinivasan2019Voder, li2023notable}, and on laying out the resulting views~\cite{qu2018keeping, shi2023reverse, dhanoa2024dtour} all operate once the analytical question has been settled, which is the step we target.

\parHeading{Navigating the space of analyses.}
EDA is often treated as open-ended pattern finding, with tools that surface variables and combinations to browse~\cite{wongsuphasawat_voyager_2016, vartak2014seedb}, but hypothesis-driven analysis is a real and distinct behavior when observed through free-form protocols~\cite{wootton2024charting}, and the two are not separable: there is no hypothesis-free exploration~\cite{felin2021data}, and exploratory and confirmatory work proceed in tandem~\cite{hullman2021designing}.
We therefore treat every state from a vague question to a grounded claim as a degree of hypothesis formation, so awareness and refinement are two faces of one thing rather than two phases of a workflow.
Visualization recommenders are a genuine precedent for reasoning over a constrained space, enumerating candidates under rules and completing them from a partial specification~\cite{wongsuphasawat_voyager_2016, 2017-voyager2, lin_dziban_2020}, but they assume the analyst can already supply that specification, which is the hard part, and they aim to navigate toward a good result rather than to convey the space itself---an aim whose evaluation criteria rarely transfer between systems~\cite{zeng2022evaluation}.
We give ranking a different job, ordering the moves we offer by how much each would distinguish the analyses downstream rather than by how good the resulting analysis is judged to be.
Multiverse analysis makes decision structure an auditable artifact~\cite{liu2020boba}, but it executes exhaustively because its outcomes are comparable point estimates; ours are heterogeneous analyses on no common axis, and the specification is still being formed.
That same precondition, a cheaply scorable objective, is what lets LLM tree search discover empirical methods automatically~\cite{aygun2025empirical}, and it does not hold when the alternatives are competing readings of an ambiguous question.
Closest in shape are systems that give guidance a representation the user can see: insight networks over findings already produced~\cite{hoffswell2025representing, li2024coinsight, aodeng2025inrectable, gathani2024groot}, in-situ next-step suggestions that leave the space implicit~\cite{chen2023whatsnext, li2023edassistant, zhao2025proactiveva}, and shared diagrams or node canvases that user and model both edit~\cite{ding2025diagram, ueno2026viscanvas}, all following Heer's argument that agency and automation are reconciled by a shared representation of possible actions~\cite{heer2019agency, kandel2011wrangler, green2014human}.
Ours differs in what that representation is made of: the intent graph is generated by a grammar, so every node is a well-formed claim and every edge a single legal refinement, which is what lets the same structure both constrain generation and be checked by the analyst.

\parHeading{Language models for open-ended analysis.}
Natural language interfaces for visualization predate large language models~\cite{gao2015datatone, narechania2020nl4dv, setlur2019inferencing} and now use them to generate and edit specifications~\cite{tian2024chartgpt, sah2024nl4dv-llm, vaithilingam2024dynavis, wang2024dataformulator, wang2025data, li2024inferring}, sometimes relaxing the turn structure itself~\cite{tabalba2023investigation, angert2023spellburst}; each returns a single view or transformation per turn rather than composing a dashboard around the question that prompted it.
A parallel thread applies language models to insight generation and summarization~\cite{weng2024insightlens, liu2024chartthinker, wang2025jupybara, lin2024inksight, zhao2024leva, ma2023insightpilot}, asking what the data shows rather than what is worth asking of it.
A third asks how far models carry \emph{open-ended} analysis~\cite{trummer2023demonstrating, guo2024talk2data, zhao2024lightva, majumder2024discovery, majumder2024data}, and benchmarks temper the answer: models give surface-level recommendations despite broad domain knowledge~\cite{gu2024blade}, cannot reliably navigate ambiguity~\cite{saparina2024ambrosia}, and produce queries that are valid but answer the wrong question~\cite{mohole2025veriminder}.
Whether such assistance helps also depends on whether the analyst can inspect, attribute, and correct it, which studies of analyst--AI collaboration find is the binding constraint~\cite{cabrera2023what, rogers2023tracing, zamfirescu-pereira2023why, vaithilingam2024imagining}.
Rather than ask a model to resolve an ambiguous question in one shot, \dageda{} makes the decomposition itself an inspectable artifact the analyst can validate, extend, and redirect, records provenance for every refinement, and derives its offers from the grammar so that an unhelpful suggestion can be traced to a rule and overridden rather than merely re-prompted.
\section{Design Considerations}
\label{sec:considerations}

Our aim is to help an analyst move from a vague question toward a set of well-formed, executable hypotheses, not necessarily a single one, along a path that is rarely linear. Two goals sit behind this aim. The analyst should be able to account for the choices the system makes on their behalf, which is a matter of deliberateness, and the process should surface alternatives the analyst would not have found alone, which is a matter of discovery. These pull in the same direction rather than against each other, because both reduce to one question: would this choice make a difference? To keep the process deliberate, every interpretive choice made on the analyst's behalf is visible, justified, and reversible, and moving between live branches costs the analyst nothing.

We derive the following design considerations from the challenges users face
during scaffolded EDA, informed by prior work on human-AI collaboration,
analysis intent, and interactive data systems.

\parHeading{DC1: Provide a Shared Representation of Analysis Intent Between the User and System.} 
A core challenge in human-AI collaboration is establishing a common 
representation of goals that both parties can inspect and reason about~\cite{heer2019agency, zhang2025exploring, hoque2024hallmark, liu2023wants}. 
Heer~\cite{heer2019agency} frames this as a shared representation of possible \emph{actions}, which presumes a space of candidate actions the system can already enumerate.
Our setting inverts that presumption: which actions are worth taking depends on an intent the analyst has not yet finished forming, so the representation has to carry intent before it can carry actions. 
A human-readable grammar provides scaffolding for the user to express their intent to the system in a structured way. A grammar also allows the user (and system) to quickly verify if LLM suggestions match the expressed intent. 
Grammars have long served as a way to abstract away execution details, letting users reason at the level of what they require rather than how the system will deliver it. 
In systems with LLM components, this role becomes more important: the grammar simultaneously steers generation by providing a structured space of valid analytical outputs, and applies a shared contract by supplying a verifiable specification against which suggestions can be checked. 
This dual function molds LLM-assisted output from opaque recommendations into auditable results, enabling users to inspect, override, and trust the system's reasoning. 

\parHeading{DC2: Verify That the Selected Dataset Is Appropriate for Answering the User's Analysis Question.} 
Users frequently approach EDA with analysis questions that are conceptually underspecified relative to the available data: the question may invoke abstract concepts that map to data variables in non-obvious ways, or it may be premised on data relationships the dataset does not actually support~\cite{passi2019problem}. 
Bridging this conceptual gap is non-trivial, as it requires anchoring abstract analytical vocabulary in domain-specific knowledge before reaching the level of concrete data variables. 
Our system addresses this through a two-stage grounding process: we first situate the user's question within an LLM-generated domain knowledge graph that captures abstract conceptual relationships, then progressively refine those concepts toward concrete data variables using structured LLM assistance. 
This approach separates the semantic interpretation of the question from its operationalization in data, reducing the risk of silently mismatching a question to an inappropriate dataset.

\parHeading{DC3: Enable Both the User and the System to Reason About the Space of All Possible Analysis Paths. }
Traditional visualization recommendation systems for EDA, predating LLMs, generate a set of ``top-$k$'' suggestions ranked and selected by manually designed metrics or ML-based models. Although these methods are designed to be data-centric, they are domain-agnostic, which misses the opportunity to incorporate domain knowledge into data exploration.  
Typical LLMs, by contrast, surface a limited set of analysis suggestions that are closely relevant to the domain context of the data, providing limited rationale for why these particular paths were selected over alternatives that were not mentioned at all~\cite{grunde2026overseeing}. 
This behavior occludes the full space of valid analyses and denies users the ability to reason about what they might be missing. 
Rather than committing to a fixed number of suggestions, our system is designed to make the entire space of valid analysis paths enumerable and navigable. 
Because the system represents all analyses derivable from the formal grammar and the knowledge graph, users can observe how it traverses this space. 
Then, the user can request explanations for analysis steps that do not make sense, intervene to redirect the system towards more relevant analysis paths, refine their original intent to better reflect their analysis goal, or accept a suggested analysis path if it matches their intended analysis goal. 

\parHeading{DC4: Enable the User to Verify How Any Given Analysis Step Is Grounded in Data Science Best Practices and the Relevant Domain Context.}
Users distrust AI-generated suggestions they cannot understand~\cite{wang2021explanations}.
In data science, trust in suggested analyses requires understanding the domain context that explains why it makes sense to combine the selected variables~\cite{gu2024blade}. Trust also requires understanding why these analysis methods are appropriate for the selected data types~\cite{gu2024blade, freund2025flowco}. 
Therefore, the system should generate explanations that facilitate the user's ability to verify its choice of domain context and analysis methods.

\parHeading{DC5: Enable the User to Compare Multiple Dashboard Designs Simultaneously. }
Because a user's analysis question may be underspecified, multiple valid analytical paths may exist for answering it, potentially yielding distinct dashboard designs that each represent a coherent interpretation of the question~\cite{lam2018bridging, battle2019characterizing, srinivasan2023bolt}. 
Prior systems that generate multiple dashboard alternatives typically do not provide users with a basis for comparing them beyond their visual appearance~\cite{dibia_data2vis_2019, luo2018deepeye, hu2019vizml}. 
We seek to provide users with the ability to compare and contrast these possible designs in terms of the line of reasoning for why the data should be analyzed along these paths to produce an answer. 
This approach is distinct from prior work, which does not ground the evaluation of multiple dashboards through comparison of sequences of analysis decisions.
This makes it possible to compare \emph{how} different interpretations of the same question lead to different analytical paths, supporting more principled selection among valid alternatives.

\parHeading{Grounding Principles.}
Five commitments follow from these considerations and constrain every part of the design.
The analyst can always see how much of the candidate space remains open, not only the current selection (\emph{space-awareness}).
Every narrowing is either a grammar consequence, auditable by construction, or an explicit inference the analyst can inspect and reverse (\emph{no silent shrinking}).
Grounding one branch never removes or hides its siblings (\emph{no premature convergence}).
The system's only license to rank, highlight, or offer a move is how much it would distinguish the analyses downstream, never raw plausibility (\emph{discriminability}).
And the shape of what is still open is shown, not only what is resolved (\emph{visible skeletons}).

\section{The Interface in Use} \label{sec:usage}

This section follows one analyst through one session to show what the grammar and the two graphs of \autoref{sec:dag-system} look like from the analyst's side of the screen.
Anna is a data analyst working for a movie production company.
Her dataset covers movies released before 2016, 3,201 rows and 16 columns spanning financial information (\eg{} budget and worldwide gross), production metadata (\eg{} genre, director, distributor, MPAA rating, and release date), and audience response (\eg{} IMDB rating and vote count).
Her brief is a single sentence: \myquote{What are the main characteristics of successful movies?}

\parHeading{How Anna does this today.}
In a commercial BI tool such as Tableau, the question is not something Anna can act on, so her first move is to make it concrete in her head.
She decides privately what \myquote{success} will mean, brainstorms conceptual variables before she looks at the schema so as not to fixate on whatever columns happen to exist, then maps those concepts onto attributes and derives the ones the dataset lacks, such as ROI or budget tier or release season.
Only then does she start building: a scatter plot for budget against gross, a bar chart of movie volume by genre, a histogram of IMDB scores, and a genre filter to tie them together.
Two costs are hidden in that sequence.
The interpretation of \myquote{success} that survives into the dashboard is the one Anna committed to before she had looked at the data, and the alternatives she considered leave no trace, so neither she nor anyone reviewing her work can tell which readings were rejected and which never came to mind.
And she learns that the dataset cannot support a reading only by building the charts that fail to answer it.

\dagTeaser

\parHeading{The interface.}
\autoref{fig:dag-teaser} shows the interface Anna works in instead.
She types her question into the question bar (A) and parses it, and the canvas (B) fills with the claims the question could resolve into, one node per candidate analysis.
Each node states the claim it makes, renders it as a sentence, and shows its slots as chips; the moves available from it hang beneath it, grouped by the state they would produce and collapsed to a count, so Anna can see how much is still open without reading through it.
Selecting a node fills the active node panel (C), which gives the claim in full, lists its slots, and, once the required ones are grounded, offers to complete the claim to a dashboard.
The knowledge graph (D) sits beside the canvas throughout, its concept and data layers joined by bridge edges and filterable by layer and edge type.
Dashboards are not a fixed region of this layout: each is completed from a particular node and opened on its own, so Anna accumulates them over the session rather than being shown a running recommendation.
The session below is otherwise entirely a matter of what she selects and what the two graphs say back.

\dagStoryboard

\parHeading{The question.}
Anna types her question and expects a dashboard.
What she gets is three readings of it side by side, each carrying the system's confidence that it is the one she meant (\autoref{fig:dag-teaser}-B, and across the top of \autoref{fig:dag-storyboard}).
The strongest, at 94\%, is a relation asking which quantitative factors drive movie success.
A comparison of success across movie categories follows at 38\%, and a trend of success over time at 22\%.
\myquote{Success} is held as an unresolved concept in all three, and no combination of columns has been enumerated under any of them.
This is the first thing Anna has to adjust to, and it is deliberate: her question admits several readings, and the system commits to none of them on her behalf.

\parHeading{The knowledge graph answers a question about the dataset.}
Before descending into any of them, Anna opens the knowledge graph to see what \myquote{success} could even mean here (\autoref{fig:dag-teaser}-D).
The concept layer places movie success at the center of a small neighborhood of readings: \textit{Box Office Success}, \textit{Audience Success}, and \textit{Critical Success}.
Following the bridge edges down to the data layer tells her which of these the dataset can carry, and on what terms.
\textit{Box Office Success} operationalizes cleanly, reaching production budget and worldwide gross.
\textit{Critical Success} does not: the only columns beneath it are rating columns, and IMDB rating in particular is an audience score standing in for a critic's judgment.
The graph says so rather than quietly substituting one for the other, and Anna decides she can work with the proxy provided she treats the two ratings as separate measures rather than as one.
That is a decision about her question, made before she has built anything, on evidence the system put in front of her rather than on charts that failed to answer it.

\parHeading{Opening a concept costs nothing.}
Anna descends \myquote{success} from the relation reading (\autoref{fig:dag-storyboard}).
It resolves into two children on the canvas, a relation over \textit{Box Office Success} and a relation over \textit{Critical Success}, and she works on both in turn.
Neither displaces the other, and the comparison and trend readings above them are untouched.
This matters more than it sounds: in her Tableau workflow, choosing one of these meant the others existed only in her memory, and revisiting one meant rebuilding it.

\parHeading{Grounding a branch to columns.}
The \textit{Box Office Success} branch descends in steps rather than in one jump (\autoref{fig:dag-storyboard}, left).
The concept first pairs with production budget, then the claim moves to production budget against worldwide gross, and only at the bottom of the chain do the slots reach the columns themselves.
At that point the node is marked \emph{renderable}: its required slots are grounded, so it can be completed to a dashboard.
Each step is a single move Anna made, and each intermediate claim stays on the canvas as its own node, so the path from an abstract concept to two concrete columns is legible afterwards rather than only in her head.

\parHeading{A commitment reshapes the claim.}
The \textit{Critical Success} branch takes a different kind of step.
Rather than descending to a column directly, Anna makes a branch selection that narrows the concept to Rotten Tomatoes rating, leaving a relation with one slot filled and one still open.
From there the branch forks.
Filling the open slot with IMDB rating produces a relation between the two ratings, which is exactly the critic-against-audience comparison the proxy warning had made her want.
Filling it with genre produces something else: genre is categorical, and because a claim's type follows from what fills its slots, the relation \emph{reshapes} into a comparison of Rotten Tomatoes rating across genre rather than forcing a category into a slot meant for a measure.
Anna reads this as the system following her rather than correcting her.
It is the one move in her session that settles what fills a slot and changes the shape of the claim in the same step.

\parHeading{Choosing without committing.}
With the comparison selected, Anna works on which categorical variable it should use.
The panel beside the canvas lists the candidates as a checklist, and ticking one re-renders the claim without changing it: the node is not replaced, nothing new appears on the canvas, and she can try major genre against creative type as often as she likes at no cost.
Only when she decides a selection is worth keeping does she commit it, which creates a child node and puts the choice on the record.
The distinction matters to her for a small reason and to the system for a larger one.
She gets to browse without littering the canvas, and the system gets a clean answer to what a node is: the claim it makes, not every variable she looked at on the way to making it.

\parHeading{Two paths, one claim.}
The comparison reading the parser ranked second, at 38\%, is not idle.
Anna descends it, and it runs the same three moves the relation reading did: success resolves to \textit{Critical Success}, a branch selection narrows that to Rotten Tomatoes rating, and a final descent settles the slot that is left.
Nothing the two readings produce along the way is shared.
They differ, too, in what that last move had to do: on the relation side it bound a category into an open slot and reshaped the claim, while on the comparison side the shape was fixed at parse time and the move only had to say which category.
They end on the same node all the same, and it carries both of them as parents.
Had the canvas been a tree it would hold two of these, one under each reading, and Anna would have no way to tell they were the same analysis.
As it is she can see that the reading the system ranked first and the reading it ranked second converge, which is a substantive thing to learn about her own question, and not something she would have found by taking the 94\% suggestion and stopping there.

\parHeading{Dashboards, and what is left open.}
From each renderable node Anna completes the claim to a dashboard, which opens on its own rather than replacing whatever she was looking at.
By the end of the session she has three open at once: production budget against worldwide gross from the box office branch, IMDB against Rotten Tomatoes rating from the critical branch, and Rotten Tomatoes rating across major genre from the node the two branches share.
Each names the template it instantiates and the bindings that fill it, and carries controls for how values are aggregated and how many categories are shown.
What she compares across them is not three chart designs but three readings of what makes a movie successful, each traceable back through the moves that produced it, and no one of them is put forward by the system as the answer.

The canvas also records what she did not do.
The moves still available from her selected node sit beside it, collapsed to counts rather than enumerated, and among them is a compose offer she never takes.
The trend reading is still where the parser left it, never descended.
These are not loose ends but the parts of the question Anna chose not to pursue, still visible at the point where she stopped.

\parHeading{What the session shows.}
Anna never told the system what \myquote{success} meant, and the system never decided for her.
Every narrowing was one she made, from a set of alternatives the system could enumerate and she could audit, and the readings she did not take are still on the canvas at the end of the session (DC1, DC3).
The one thing she settled before building anything came from the knowledge graph reporting the terms on which the dataset could operationalize a concept (DC2), and the convergence she found was legible only because each path is recorded on the node it reached (DC4).
What she leaves with is not a dashboard but several, positioned against each other by the reasoning that produced them (DC5).
One thing the session does not show is the return path: \dageda{} does not yet read Anna's interactions with a rendered dashboard back into the intent graph, which we take up in \autoref{sec:dag-future}.

\section{System Design}
\label{sec:dag-system}
\label{sec:implementation}

\dagArch

\dageda{} turns an ambiguous analytical question into a navigable space of concrete analyses. A session starts when the analyst types a question over a dataset, such as \myquote{What are the main characteristics of successful movies?} From the question and the dataset schema, the system builds two structures (\autoref{fig:dag-arch}). A knowledge graph (\autoref{fig:dag-arch}, \textcircled{\footnotesize B}) grounds the vocabulary of the question in the data, connecting the abstract terms the analyst reasons with to the columns the dataset actually provides. An intent graph (\autoref{fig:dag-arch}, \textcircled{\footnotesize A})  lays out the analyses the question could resolve into, one node per candidate analysis. Every node is a \emph{partial analysis}: a claim with some of its parts settled and the rest still open. The analyst refines the question by moving through the intent graph one step at a time, settling what an ambiguous term denotes, committing to how two variables relate, drilling into a mediator, or binding a slot to a column. Each move is a single grammar rule made interactive, and each fully specified analysis maps to a recommended dashboard.

The one idea that holds the graph together is that a node is identified by the claim it makes, not by the path the analyst took to reach it. Two routes that arrive at the same claim over the same columns arrive at the \emph{same} node, so the intent graph is a DAG by construction rather than a tree the system has to prune. This is what makes exploration cost-free. Branches never overwrite one another, a node the analyst leaves is exactly where they left it when they return, and backing out of a step is just stepping back to a claim that was never destroyed.

We organize this section around the design considerations from Section~\ref{sec:considerations}. A grammar of analytical claims gives the user and the system a shared representation of intent (DC1). The knowledge graph and its two-stage grounding test whether the dataset can answer the question (DC2). The intent graph and its one-step offers keep the space of valid analyses enumerable and navigable rather than reducing it to a ranked shortlist (DC3). Provenance and inspectable candidates let the analyst verify why each step was taken (DC4). Convergence and side-by-side comparison let the analyst hold competing interpretations of the same question against each other (DC5). We describe the grammar, then the two graphs it works over, then how they fit together, giving each its concrete realization where it is introduced.

\subsection{The Grammar}
\label{sec:dag-grammar}

The analyst and the system work over one shared representation of intent: a small grammar of analytical claims (DC1). We describe it in the large before pinning down the small structure that stands in for a claim.

A claim is drawn from a small typology of shapes~\cite{choi2019concept}: a \emph{value} summarizes one variable, a \emph{trend} follows a variable over time, a \emph{relationship} ties two variables together, and a \emph{comparison} contrasts a measure across a category. A relationship folds correlation and causation into one shape, committing a direction only once the evidence supports it rather than assuming one at the outset. Each claim is built from named \emph{slots}, each carrying a role and a type (measurable, categorical, or temporal). A claim's type follows from what fills its slots rather than a label the analyst sets: a lone measure reads as a value, a measure and a category as a comparison, two measures as a relationship, a measure over time as a trend. Because the type follows the slots, adding or changing a slot can reshape the claim, so naming a categorical driver of an outcome turns a relationship into a comparison across that category rather than forcing the category into a measure. A slot may be an open placeholder, an unresolved concept (\myquote{success} before it is pinned to revenue or ratings), or grounded to one or more columns. \emph{Grounding} is the act of binding a slot to data; a claim becomes executable, and offers a dashboard, once its required slots are grounded. When a concept has no adequate column, the system says so and holds it anyway, which is itself an answer: the dataset does not support that reading of the question (DC2).

The analyst refines a claim along two independent axes, which the draft grammar had conflated. \emph{Referential} moves resolve what a term denotes, opening an unresolved concept into the concepts it could mean. \emph{Relational} moves change the claim's structure, either the shape it instantiates or how it composes with other claims. The two constrain each other: fixing \myquote{success} to a categorical genre rules out the relationships that would treat it as a measurable outcome, and asking for a driver of an outcome requires that outcome to be measurable. Keeping the axes separate lets a commitment on one prune the other, and lets the analyst make the two kinds of commitment independently rather than through a single tangled choice.

Formally, a claim is a \emph{state}: a set of at most two typed slots $S = \{(\tau_1, b_1), \dots, (\tau_k, b_k)\}$ with $k \le 2$. Each type $\tau_i$ is measurable, categorical, or temporal ($Q$, $C$, or $T$), and each binding $b_i$ is an open placeholder, a concept in the hierarchy, or a set of grounded columns. The type multiset alone decides the claim. A pure function $f$ maps it to a hypothesis type ($f(\{Q\}) = \text{Value}$, $f(\{Q,Q\}) = \text{Relation}$, $f(\{Q,C\}) = \text{Comparison}$, $f(\{Q,T\}) = \text{Trend}$), and a second function $g$ maps a hypothesis type to the dashboard template or templates that render it. Neither $f$ nor $g$ is stored on the node; both are computed on demand, so the type and the template of a state are always a function of its slots and cannot drift out of sync with them. A type multiset with no image under $f$ is a partial state: a legitimate stopping point the analyst can extend or compose, not an error.

\dagGrammarTable

We separate the engine from the typology: the engine instantiates states, discovers candidates, and applies productions, while the typology says which claims exist and how they render. The typology is realized as the tables behind $f$ and $g$ rather than as a document the engine interprets, so swapping those tables swaps the analytical vocabulary without touching the engine, and the Choi~\etal{} set is the reference instantiation. \autoref{tab:dag-grammar} gives that instantiation, and it is short: $f$ is defined only up to arity two, so a base state binds at most two variables and every analysis over three or more is a composite. The bound is not a budget on complexity but a consequence of what an unordered multiset can say. Two variables need no designation, because their types settle their roles: in $\{Q,C\}$ the measure is what is compared and the category is what it is compared across, and no other reading is available. A third variable always introduces a choice the multiset cannot record. Three measures do not say which is the mediator; a measure with two categories does not say which category the other is read across. Composition exists to supply exactly that missing designation, which is why the boundary falls where it does rather than at four or five slots. One consequence matters downstream: an open slot's type is provisional and adopts the type of whatever it first binds to. Binding a categorical driver to an open slot therefore reshapes a relationship into a comparison, rather than pinning a category into a measurable slot.

This is what makes the system's behavior auditable (DC1). Every candidate the system surfaces, every move it offers, and every check on whether an analysis is well-formed follows a grammar rule rather than ad hoc logic, so the analyst can hold a suggestion to the same rule the system does.

\subsection{The Intent Graph}
\label{sec:dag-intent-graph}

The intent graph (\autoref{fig:dag-arch}, \textcircled{\footnotesize A}) records where the analyst has been and where they can go. Each node is a claim, an instance of the grammar, and each edge is a refinement move. The graph is a DAG rather than a tree because two paths often reach the same claim: an analyst can start from \myquote{What drives revenue?} and drill down, or ground revenue and budget separately and compose them, and both routes land on the same relationship over the same two columns. Because a node is identified by the claim it makes, not by the path taken to reach it, the two routes do not produce two nodes to be reconciled after the fact; they produce one node with two parents. There is nothing to merge, because the node was never two things. How each path reached the node is kept as provenance, so the analyst can still see why the node is on their path.

Because paths converge (see \autoref{fig:dag-storyboard}), the order in which analyses appear is not a fixed sequence. The shortest route from the root question to any node the analyst has not yet visited is recomputed as they descend, and moving sideways between live branches costs nothing: the state of a branch is preserved when the analyst leaves it and restored when they return. This is what lets the analyst hold several readings of a question open at once instead of committing to one and backtracking (DC3).

The system realizes this identity by \emph{content-addressing}. A node is addressed by what it is, not by how it was reached: the system computes a key from the state itself, a normalized hash of its slots and bindings, and that key is the node's identity, exactly as a version-control system names an object by a hash of its content so that identical content is one object no matter who produced it. Interning a newly produced state either returns the existing node with that key or creates a new one, and records the move as a provenance edge. Content-addressing collapses a cluster of algorithms an earlier design carried separately. Consolidation is not a pass that finds and merges duplicates: two paths that produce the same state produce the same key and therefore the same node, with several incoming edges. Reversibility is not a snapshot the system restores: a move only ever adds a node, so backing out is re-selecting the parent, which was never altered. Lateral movement between branches is free for the same reason, since a state is immutable once interned and nothing one branch does can reach into another. The provenance graph is kept acyclic: a move that lands on an existing ancestor is recorded as a revisit rather than a back-edge, so the history stays a legible record of productions rather than a tangle.

The edges of the graph are \emph{productions}, and they come in a few families. \emph{Structural} productions reshape the type multiset: add a slot, drop a slot, or retype one, each of which changes the hypothesis type and hence the template. \emph{Binding} productions move a single slot through the concept hierarchy: descend to a more specific concept or column, or ascend to a more general one. A descent that reaches a column grounds the slot, and committing a hand-picked set of columns is a binding production of its own. \emph{Composition} productions create or unwind the composite structures of Section~\ref{sec:impl-composition}. Every production validates its precondition (arity at most two, no two slots grounded to the same columns, a descent staying within its slot's subtree), and an illegal move is refused rather than allowed to produce an invalid state. The offers the analyst sees at a node are its \emph{frontier}: the set of productions found legal from the current state, computed on demand rather than stored as ghost nodes. Each frontier entry carries the state it would produce, and therefore the hypothesis type the analyst would get, together with a confidence score drawn from the knowledge graph, a description of the resulting claim, and a fixed goal explaining why the analyst might want it. A separate preview walks the frontier several steps ahead, following the highest-confidence child at each level, so the analyst can see where a line of refinement leads before committing to it; nothing along the preview is materialized until it is chosen. Selecting and toggling columns within a slot to watch the effect is view-state on the node, not a production: it changes what is displayed, not the claim's identity, and only the explicit commit of a chosen set enters identity and creates a child.

\subsection{The Knowledge Graph}
\label{sec:dag-kg}

The knowledge graph (\autoref{fig:dag-arch}, \textcircled{\footnotesize B}) bridges the gap between the concepts the analyst reasons with and the columns the dataset stores. It has two layers. The \emph{concept layer} (\autoref{fig:dag-arch}, \textcircled{\footnotesize C}) holds the domain concepts implied by the question and how they relate, so \myquote{success} sits near \myquote{financial performance} and \myquote{audience reception}. The \emph{data layer} (\autoref{fig:dag-arch}, \textcircled{\footnotesize D}) holds the dataset's columns and the causal or associative relationships among them. The two layers are joined by \emph{bridge edges} that operationalize a concept as one or more columns, each tagged with whether the column is available, a proxy, or missing. Bridge edges are connections, not a third layer of their own. The genuine third structure is the intent graph itself, which attaches to concepts and columns as the analyst grounds slots over the course of the session.

Splitting concepts from data lets the system reason about the question before it reasons about the dataset. It first situates the question in the concept layer, then follows bridge edges to find the columns that operationalize each concept. When a concept operationalizes only through a proxy or has no column at all, the availability tag surfaces the mismatch instead of silently dropping the concept.

The graph is built before exploration begins, in three LLM passes from the question and the schema (\autoref{fig:dag-arch}, \textcircled{\footnotesize 2}). The first pass generates the concept layer, the domain concepts the question implies and the relationships among them. The second pass grounds each concept to columns, routing through a canonical data concept so the match is made against a domain standard before it is made against a particular column name. The third pass adds the causal and associative edges among columns that constitute the data layer, together with the bridge edges that operationalize each concept, tagged with whether the column is available, a proxy, or missing. The two layers are cached on different keys, because they depend on different things. The data layer describes relationships that hold among the dataset's columns whatever is being asked of them, so it is keyed on the dataset alone and is reused across every question posed against it. The concept layer and the bridge edges depend on the question as well, so they are keyed on the question and the dataset together; returning to a question reuses them, and posing a new question over a dataset already seen rebuilds only this question-specific part on top of the data layer already in hand. Manual edits to the graph are versioned and can be rolled back, and are replayed when the graph is rebuilt, so the analyst's corrections survive. The concepts and bridges the graph produces are the seed for the hierarchy the analyst then refines slot by slot.

\subsection{How the Two Graphs Work Together}
\label{sec:dag-integration}

The grammar says what claims are possible; the two graphs make a claim answerable, and they divide the labor between them. The knowledge graph carries meaning: what the concepts in the question are, and which columns could operationalize them. The intent graph carries exploration: the claims the analyst has formed and the moves between them. Neither is useful alone. A claim in the intent graph is only executable once its slots are grounded, and grounding is exactly a walk through the knowledge graph from a concept to the columns beneath it, drawn as the \emph{binding} edge of \autoref{fig:dag-arch}.

The two meet at every move. When the analyst refines a claim, the intent graph asks what could fill a slot, and the knowledge graph answers by following bridge edges from the slot's concept to the columns that operationalize it, surfacing those as candidates. When a slot is still open beside one that is already grounded, the knowledge graph's causal edges seed it, so asking for a driver of revenue starts from revenue's actual factors rather than from every column in the dataset. And when the analyst grounds a slot, the claim advances in the intent graph and its offers are recomputed, which may send the next question back to the knowledge graph. Grounding is the loop that binds the two: the intent graph poses the questions, and the knowledge graph answers them in data.

Four components run this loop. A \emph{parser} (\autoref{fig:dag-arch}, \textcircled{\footnotesize 1}) reads the question and proposes one or more claims, each with hints for the slots the question mentions. A \emph{knowledge-graph builder} (\autoref{fig:dag-arch}, \textcircled{\footnotesize 2}) constructs the concept and data layers and the bridges between them. A \emph{propagator} maintains the intent graph: it applies the analyst's moves, consults the knowledge graph to discover and rank the candidates that fill a slot, and records where paths converge. A \emph{canvas} (\autoref{fig:dag-teaser}) renders both graphs and the dashboards, and routes the analyst's moves back to the propagator. The two components that need world knowledge, the parser and the construction of the knowledge graph, sit at the edges of the pipeline; everything between them, the enumeration of claims and the checking of what is valid, follows grammar rules and runs deterministically. This keeps the parts of the system the analyst navigates predictable and auditable (DC4), and confines generation to the two points where the system genuinely needs world knowledge. Two stages of the loop rely on generated knowledge and warrant a closer look: turning the question into claims, and filling a slot from the hierarchy.

\parHeading{Parsing.} Parsing (\autoref{fig:dag-arch}, \textcircled{\footnotesize 1}) is the one place the system reads the analyst's free text. The parser sends the question and the dataset schema to an LLM and receives one or more interpretations, each a state with hints for the slots the question mentions. A question that admits several readings produces several interpretations, which become parallel framings in the intent graph rather than a single classification the analyst has to accept. What the parser returns is a partial state: some slots hinted and the rest left open. Turning those hints into slot bindings is deterministic and needs no further LLM call. For each slot the system decides, from the hint alone, whether the question bound it to a single concept, named an open concept with several readings, or said nothing about it. When the type multiset the parser produces has no image under $f$, the reading is lifted into a well-formed composite, so that every entry point is a valid state or composite: a categorical mention alongside a relationship, for instance, becomes a moderation.

\parHeading{Discovery and the dynamic hierarchy.} Discovery fills a slot by walking the concept hierarchy to the columns beneath a concept, following bridge and part-of edges until it reaches data, which it surfaces as checkbox candidates rather than as new nodes, so a deep operationalization does not spawn a subtree. The hierarchy is generated lazily and conditioned on context rather than precomputed once. A concept's children are produced when the analyst first descends into it, and an open slot beside a grounded one is seeded from the grounded variable's causal neighborhood in the knowledge graph, so asking for a driver of revenue starts from revenue's actual factors, of every type. Because a categorical factor keeps its type, descending into one reshapes the claim from a relationship into a comparison rather than forcing it into a measure. Two obligations come with generating the hierarchy on the fly. First, \emph{deduplication}: a generated concept that already exists must resolve to the existing node, or content-addressing silently breaks. A concept carrying a knowledge-graph identifier is resolved by that identifier; a free concept falls to a normalized-label match and, if still ambiguous, an LLM check for semantic equivalence. Second, \emph{bounded generation}: every generation call is capped in depth and breadth, so the termination guarantees the rest of the system relies on hold against a finite hierarchy. When a concept bottoms out at the cap without reaching data, the system grounds to the nearest columns rather than dead-ending. When a column is semantically apt but the wrong type for a slot, the system asks the LLM for a transformation, a binning or a ratio, and materializes the result as a derived column.

\subsection{Composition}
\label{sec:impl-composition}

\emph{Composition} reaches beyond a single filled-in claim, building an analysis from several claims when one is not enough. \emph{Moderation} facets a relationship by a category, asking whether the budget--revenue relationship holds across genres; \emph{mediation} places a variable on the path between two others, asking whether budget drives revenue directly or through marketing spend; and \emph{comparison} sets two full analyses side by side, comparing not two columns but two lines of reasoning. Composition is why the base claims can stay small: rather than growing a claim to three or more slots, the analyst composes bounded pieces. The division of labour is sharp because the arity bound of \autoref{sec:dag-grammar} makes it so. A base state carries variables whose roles their types already settle; a composite carries a designation that no multiset can express, naming the mediator of a chain, the variable a pattern is conditioned on, or the axis two analyses are aligned along.

A composition is a recursive extension of the state. Writing $\textit{Spec}$ for either a base state or a composite,
\[
    \textit{Spec} \;::=\; \textit{State}
    \;\mid\; \text{Condition}(\textit{Spec}, z)
    \;\mid\; \text{Chain}(\textit{Spec}, \textit{Spec}, m)
    \;\mid\; \text{Compare}(\textit{Spec}, \dots).
\]
\dagCompositeTable
\autoref{tab:dag-composites} gives each form its well-formedness condition and the layout it compiles to.
In the grammar, moderation, mediation, and comparison are $\text{Condition}$, $\text{Chain}$, and $\text{Compare}$ respectively, which lifts them out of the flat templates of earlier drafts: a chain is literally two relationships that share a column, not a bespoke mediation template. Well-formedness here is \emph{relational}, a property of how the sub-specs fit together that a single state cannot express. A $\text{Compare}$ requires its legs to share a hypothesis type and to be distinct; a $\text{Chain}$ requires both legs to be relationships that actually share the mediator; a $\text{Condition}$'s conditioning variable must be categorical, and may be left open to be bound afterward. Each composite compiles to a qualitative layout rather than a single template: faceted small multiples for a condition, linked panels for a chain, overlaid or side-by-side panels for a comparison. Creation productions bridge from a base state to a composite, and because $f$ stops at arity two they are offered wherever a third variable is added at all: from a relationship, adding a category offers a moderation and adding a mediating measure offers a chain; from a comparison, adding a second category offers a moderation; from a trend, adding a second measure offers a comparison of trends. There is no competing route, since the multiset the added slot would produce has no image under $f$ and is never surfaced as a state. Inverse productions unwind a composite back toward the base.

\subsection{Implementation Notes}
\label{sec:impl-interface}

The canvas is a Next.js and React application. It lays out the intent graph with ReactFlow and the concept trees with D3, and holds the analyst's selections, folds, and open dashboards, including the fluid, uncommitted column selections of Section~\ref{sec:dag-intent-graph}, in a client-side store. A concept's variable leaves are grounded through a checklist that binds the checked set in one action, and a node-level group-by picker facets that node's dashboards over categorical columns at render time. Each node carries a natural-language description rendered from its state and specialized by how much of it is grounded, with the slots rendered as chips the analyst can click to select the corresponding data.

Dashboards are pre-authored Mosaic~\cite{heer2024mosaic} template specifications, compiled to SQL and run in the browser over DuckDB-WASM~\cite{raasveldt_duckdb_2019}. The feasible templates for a terminal node are filtered and ranked by how well its bindings cover each template's roles and by the availability of the data, with the best marked as recommended, and a node can open several dashboards to branch into more than one design. The mapping from a grounded claim to its dashboards is itself a small graph rather than a single choice.

\section{User Study Design}
\label{sec:dag-eval}

We conducted a user study to evaluate how \dageda{} supports open-ended exploratory data analysis through its two core components, the intent graph and the knowledge graph.
Rather than measuring task completion against a ground truth, which open-ended EDA does not admit, we study how analysts interact with the two structures while exploring an unfamiliar dataset to answer a question of their own.
Our overarching concern is whether the system \emph{augments} analytical reasoning rather than automating or confusing it.
We organize the study around four questions, which we label \textbf{SQ1}--\textbf{SQ4}. 

\parHeading{SQ1: Intent graph alignment and user agency.}
Does the system's decomposition of a question reflect the direction the analyst is actually pursuing, and does moving through the nodes help them form their intent?
Are the alternative analyses the system surfaces inspiring and useful, or noise?
We are particularly interested in how analysts navigate between following the system's suggested scaffolding and redirecting the system with their own reasoning, which speaks directly to DC1 and DC3.

\parHeading{SQ2: Knowledge graph utility.}
Does the knowledge graph productively inform and reframe the analyst's thinking behind their question?
This tests whether externalizing the LLM's domain knowledge into an inspectable structure (DC2, DC4) changes how analysts interpret both the question and the dataset, or whether it is treated as decoration.

\parHeading{SQ3: Scaffold or menu, across domains.}
Are the intent and knowledge graphs used as scaffolds for the analyst's own thinking, or as a menu of pre-set options to pick from?
Because participants formulate their own question on both datasets, we can observe this stance twice per participant, once in an everyday domain and once in a specialized one, and ask whether analysts lean harder on the system's suggestions when they have less domain knowledge of their own to bring.

\parHeading{SQ4: Cognitive load and perceived performance.}
How much effort does it take to work with the system relative to what analysts feel they accomplish?
A scaffold that is faithful but expensive to operate is not a usable scaffold.


\subsection{Participants}
\label{sec:dag-eval-participants}

We recruit 10--12 participants in two phases.
First, we distribute a short screening survey (approximately 5--10 minutes) through university mailing lists, research lab channels, and professional networks.
The survey asks about the respondent's role, domain, and years of experience, how often they build dashboards, how and how much they currently use LLMs in their data work along with the biggest challenge they face in doing so, what other tools they use, and how they decide where to start a new analysis; an optional second section collects age group, gender, and education.
Beyond establishing eligibility, these responses give us a baseline account of each participant's current practice against which to read their behavior with \dageda{}.

We target data analysts, data scientists, business intelligence professionals, and similar practitioners with considerable experience creating dashboards or conducting exploratory data analysis, in either an academic or a professional context.
Participants must be 18 or older, and we recruit across a variety of domains so that the questions participants formulate in the tasks are not drawn from a single analytical culture.
Eligible respondents are invited to a 90-minute session and compensated with a \$45 gift card.

\subsection{Setup}
\label{sec:dag-eval-setup}

All sessions are conducted remotely over video conferencing, with the participant driving \dageda{} in their own browser through screen sharing.
Before each session we confirm that the recording software and interaction logging are functioning, that the system and both task datasets are loaded and ready to share, and that the tutorial video and warm-up materials are prepared.
We also prepare one reference sheet per dataset, each containing a short summary of the dataset followed by a list of its attributes with a brief description and unit of measurement for each, and a slide of reminders for the participant to keep visible during the tasks.

With the participant's consent we collect three kinds of data.
First, \emph{audio and screen recordings} of the full session, which capture the think-aloud narration alongside what the participant is doing in the interface.
Second, \emph{post-session questionnaire and interview responses}.
Third, \emph{system interaction logs} captured throughout both tasks, which record path and branch creation in the intent graph, brushing on charts, highlighting and dragging within the knowledge graph, and dashboard bookmarking.
The logs let us reconstruct each participant's trajectory through the intent graph after the fact and align it with the moments in the recording where they explained what they were doing, so that a claim about how the scaffolding was used can be grounded in the actual sequence of moves rather than in recollection alone.
Recordings and logs are used for research purposes only and are not shared outside the research team.

\subsection{Procedure}
\label{sec:dag-eval-procedure}

Each session lasts approximately 90 minutes.
We open by describing the session structure, asking the participant to think aloud throughout, and obtaining consent to record after stating exactly what the recordings and interaction logs capture.
A short pre-recorded video walkthrough of the interface follows, which keeps the introduction identical across participants.
Participants then spend five minutes warming up on the movies dataset with a question we supply, \myquote{what makes a movie successful}.
The warm-up serves two purposes: it separates the cost of learning the interface from the cost of reasoning with it, so that the first task does not conflate the two, and it demonstrates what a guiding question looks like at the level of abstraction \dageda{} is designed for, one that names a concept the dataset does not directly contain, which participants need before writing their own.

Participants then complete two 25-minute tasks, separated by a two-minute break, one on the movies dataset and one on NHANES, the National Health and Nutrition Examination Survey.
At the start of each we hand the participant a reference sheet summarizing the dataset and describing each attribute with its unit of measurement, and ask them to formulate their own high-level question and state it aloud before the timer starts, along with what they are looking for in it.
This gives us a record of their intent that is independent of anything the system later shows them, and against which the system's decomposition can be compared.
The task is to explore the data to answer that question, not to produce a dashboard: dashboards are what the system offers at the analyses the participant reaches, and whether they open, compare, or bookmark any is itself an observation rather than a requirement.
Participants may refine their question as they go, which is deliberate, since an intent graph that never prompts a revision would be scaffolding a fixed target rather than a reasoning process, and refinement is itself an observable outcome.
The order is fixed rather than counterbalanced, so that the movies task follows directly from the warm-up on the same data and the specialized domain comes second.
We give a two-minute warning in each task and afterwards ask whether there is anything the participant would like to note about what they were trying to do.
The experimenter is present throughout but does not guide the analysis; when a participant is stuck, we ask them to narrate their thinking and keep interacting with the system rather than intervene.

Immediately after the second task, participants complete a five-minute questionnaire while the experience is still fresh.
Its first half adapts the NASA-TLX~\cite{hart1988nasatlx} dimensions to the analytical setting, asking how mentally demanding and time-pressured the tasks felt, how much effort went into working out how to operate the system, how successfully participants feel they explored their question, and how frustrating the system's unexpected responses were (SQ4).
Its second half targets the two components: whether being redirected along the system's suggestions was helpful or disruptive, whether the intent graph felt like a scaffold for the participant's own thinking or a menu of pre-set options, how much the knowledge graph improved their understanding of the dataset, and, as free response, whether what they arrived at answers the question they started from (SQ1--SQ3).

A ten-minute semi-structured interview closes the session.
We ask participants to recall specific moments where the system's decomposition of the question matched or diverged from what they were trying to explore (SQ1) and where its interpretation of a concept felt accurate or surprising (SQ2), which features were most useful and which most confusing, and what they would change about how the intent graph responds to their actions.
Two paired questions ask for a moment where the system helped their reasoning and a moment where it got in the way; we keep them symmetric so that participants are as licensed to report breakdowns as successes.

\section{Discussion}
\label{sec:dag-discussion}

\subsection{Implications}

\parHeading{A grammar is what makes an LLM's contribution checkable.}
The role the grammar plays here is not primarily to define a language but to bound one.
Because every node the system can offer is a well-formed claim and every edge a single legal refinement, an offer that is wrong is wrong in a way the analyst can name: it binds the wrong column to a slot, or refines along a relation the question did not call for.
This is a different footing from a system whose suggestions arrive as free-form text, where an unhelpful suggestion can only be re-prompted because there is no structure against which to say what was unhelpful about it.
The general implication is that steering generation and auditing it are the same problem: a representation expressive enough to constrain what a model may produce is, by that fact, expressive enough for a person to check what it did produce.

\parHeading{Revisability is the right response to unverifiable knowledge.}
\dageda{} depends on a language model for exactly the knowledge that cannot be checked against the data.
That two rating systems aggregate different voter pools, or that a metric is a proxy rather than a measure, is domain knowledge with no oracle in the dataset to verify it against, and no amount of grounding in the schema will settle it.
The resolution is not to verify but to make revisable, which is why the knowledge graph is an editable object rather than a fixed input: the analyst, not the system, is the authority on their own domain.
Where a claim can be checked it should be; where it cannot, exposing it for correction is not a weaker substitute but the appropriate response to a different kind of claim.

\parHeading{Ranking can serve navigation instead of standing in for judgment.}
Recommenders order candidates by how good the resulting analysis is estimated to be, which presumes a scoring function over analyses.
No such function is available for competing readings of an ambiguous question, since the alternatives are interpretations rather than implementations and none of them is correct in a sense a metric could capture.
\dageda{} responds by changing what ordering is for: offers are ordered by how much each move would distinguish the analyses downstream, a property of the space rather than a verdict on the analysis.
This suggests a general move available to systems operating where quality is undefined, namely to rank for the information a choice yields rather than for the value of its outcome.

\subsection{Limitations and Future Work}
\label{sec:dag-future}

\dageda{} has not been evaluated with analysts, so the claims of this chapter rest on the design and its rationale and on the walkthrough of \autoref{sec:usage}, not on evidence about how analysts behave when using the system.
The questions this leaves open are the ones the design cannot settle on its own: whether the intent graph's progression aligns with the direction an analyst is actually pursuing, whether the knowledge graph meaningfully augments their understanding of the data domain and the intent graph branching, and whether the two graphs are used as scaffolds for the analyst's own reasoning.
Answering them calls for a study in which practitioners explore unfamiliar datasets under questions they formulate themselves, which we regard as the immediate next step for this work.

Two further limitations follow from what the system is made of.
The knowledge graph is generated by an LLM from prompts centered on the schema and the question, so its coverage of a domain is bounded by what the model knows and biased toward domains well represented in its training data; a specialized field with idiosyncratic vocabulary is the case most likely to produce a plausible-looking graph that a practitioner would not recognize.
Editability is the mitigation, and it shifts effort onto the analyst rather than removing the problem.
Separately, the guarantee that the space of valid analyses stays enumerable and navigable is a guarantee about the representation, not about the display: the number of derivable claims grows quickly with the size of the schema and the richness of the knowledge graph, and while the interface offers one legal step at a time rather than the whole space at once, we have not established at what scale the graph stops being legible to a person.

A grounded state is a specification, not a chart. Its slots, types, and bindings describe an analysis precisely enough to drive targets other than a dashboard, and we see two directions worth pursuing. A terminal relationship whose outcome and predictors are grounded could drive a notebook regression, with the bindings surfaced as editable variables; composable notebook cells in the style of Engraft~\cite{engraft2023horowitz} would let the terminals of a path merge into one environment whose shape mirrors the analyst's traversal. A terminal could also bound a Boba~\cite{liu2020boba} multiverse: the grounded slots fix the decisions in scope, and the candidates the analyst ruled out during disambiguation give a grammar-derived account of what the multiverse need not enumerate, a boundary that multiverse tools otherwise ask the analyst to draw by hand.

Two further directions extend the reasoning rather than the output. The grammar of Suh~\etal~\cite{suh2022grammar} raises the possibility of measuring and comparing hypothesis spaces by coverage and containment, which it does not itself realize in a system; our intent graph is a natural place to compute such measures over the space an analyst has opened. And the discriminability of options, which we approximate here through relevance and availability, could be computed directly as the divergence in what each candidate unlocks downstream in the graph, turning the principle into the system's ranking signal.

A final direction concerns the signals the system learns from. Today the analyst's influence on ranking is entirely constructive: they build claims up, and the system orders what it offers by relevance in the knowledge graph and by data availability, never by what the analyst has turned away from. Treating the analyst's moves themselves as \emph{feedback signals} would let that ranking, ideally the discriminability signal of Section~\ref{sec:considerations}, adapt to the analyst and not to the graph alone. The clearest \emph{positive} signal is an evidence-backed commitment: when the analyst weighs several candidate columns for a slot against the data and then commits to one (an interaction we sketch but set aside here), the choice reports what they find worth pursuing, not merely that the option was well-formed, and read as a signal it could lift kindred concepts and columns across the branches still open.

The signal the current design lacks entirely is a \emph{negative} one: letting the analyst deliberately remove the paths they judge irrelevant and reading those removals as evidence of what not to pursue. Removal carries information precisely because exploration is otherwise non-destructive. Content-addressing keeps every sibling alive, and nothing needs pruning for the graph to stay correct (Section~\ref{sec:dag-intent-graph}), so an explicit deletion is unambiguous: the analyst is not tidying up but stating disinterest. The system could then down-weight directions like the one removed, de-prioritizing the concepts, columns, and claim shapes it shared, so discarding a \myquote{success as awards} reading pushes award-related offers down across the branches that remain. Two constraints keep this honest, and are what make feedback a design problem in its own right rather than a single new move: a demoted direction must show that it was demoted because the analyst pruned a sibling, so the space never shrinks silently (Section~\ref{sec:considerations}), and a removal must be as reversible as any other move, so one the analyst reconsiders can be undone without the subtree having been lost. How far a removal should generalize, and how strongly it should propagate, are the open questions.


\begin{acks}
\end{acks}

\bibliographystyle{ACM-Reference-Format}
\bibliography{sections/bibliography}


\end{document}